\documentclass[twocolumn,amsmath,aps,amssymb,superscriptaddress]{revtex4}
\usepackage{graphicx}
\usepackage{dcolumn}
\usepackage{setspace}
\usepackage{epstopdf}
\begin{document}
\title{Theory of possible effects of the Allee phenomenon on \\refugia
       of the Hantavirus epidemic}

\author{ Niraj Kumar}
\affiliation{Consortium of the Americas for Interdisciplinary Science and
         Department of Physics and Astronomy, University of New Mexico, Albuquerque,
         New Mexico 87131, USA}
\author{M. N. Kuperman}
\affiliation{Consortium of the Americas for Interdisciplinary Science and
         Department of Physics and Astronomy, University of New Mexico, Albuquerque,
         New Mexico 87131, USA}

\affiliation{Centro At{\'o}mico
Bariloche and Instituto Balseiro, 8400 S. C. de Bariloche,
Argentina\\
Consejo Nacional para las Investigaciones
Cient\'{\i}ficas y T\'ecnicas, Argentina}
\author{V. M. Kenkre }
\affiliation{Consortium of the Americas for Interdisciplinary Science and
         Department of Physics and Astronomy, University of New Mexico, Albuquerque,
         New Mexico 87131, USA}
\begin{abstract}
We investigate possible effects of high order nonlinearities on the shapes of infection refugia
of the Hantavirus epidemic.
We replace Fisher-like equations that have been recently used to
describe Hantavirus spread in mouse populations by generalizations
capable of describing Allee effects that are a consequence of the high order nonlinearities. We analyze the equations to
calculate steady state solutions. We study the stability of those solutions under
physical conditions and compare to the earlier Fisher-like case.
We  consider spatial modulation of the environment and find that unexpected results appear, including a bifurcation that has not been studied before.
\end{abstract}
\maketitle{}

\section{Introduction}
Techniques of nonlinear physics and mathematics are finding much application in recent times in biological and ecological systems, which, in turn, are enriching nonlinear science. One example is the spread of epidemics \cite{MURR,LI,COSNER,Levin,KOT}, in particular the Hantavirus \cite{YATES} that has been recently modeled \cite{AK,KP,KPHA,KLAC}
with the help of the Fisher equation \cite{FISHER} with internal states representing
infection or its absence in the mouse population. Our purpose in this
paper is to study the consequences of Allee mechanisms (to be described below) in the
dynamics of mice that are the carriers of the Hantavirus. Without these effects, the (Fisher-like) equations
are \cite{AK,KP},
\begin{eqnarray}{\label{F_e1}}
  \frac{\partial M_S}{dT}&=&bM-cM_S-\frac{M_SM}{K}-aM_SM_I+D\frac {\partial^2M_S}
  {\partial X^2}, \nonumber \\
  \frac{\partial M_I}{dT}&=&-cM_I-\frac{M_IM}{K}+aM_SM_I
    +D\frac{\partial^2M_I}{\partial X^2},\nonumber\\
  \frac{\partial M}{\partial T}&=&bM-cM-\frac{M^2}{K}
   +D\frac{\partial^2M}{\partial X^2},
\end{eqnarray}
where we display a $1-d$ system for simplicity. Here, $M_S$ and $M_I$ are,
respectively, densities of susceptible and infected mice and
$M=M_S+M_I$ is the total mouse density, $a$ controls transmission
of infection on encounter, $b$ and $c$ are rates of birth and death respectively and
$D$ is the mouse diffusion coefficient. The third of the Eqs.(\ref {F_e1}) is obtained
simply by adding the first two and is of the standard Fisher form \cite{MURR}. The
first two possess the characteristic features of the Hantavirus \cite {YATES} that
mice are never born infected and that they are unaffected in any other way (for instance,
they do not die faster) when infected. The logistic reaction term in the last of the
Eqs.(\ref{F_e1}) is made up of a linear growth term $(b-c)M$ and bilinear depletion
term $-M^2/K$. Such a description of population dynamics is widespread \cite{MURR}.
The steady state homogeneous solutions for
$M$ are \cite{AK}: $0, K(b-c)$. The first of these is unstable whereas the second, $M=K(b-c)$
is stable.\\

The Fisher analysis of the Hantavirus is based on the use of
a logistic reaction term, whereas the investigation
we report in the following assumes a Nagumo term which provides an additional zero in the
nonlinearity relative to the logistic case. The physical content behind
such a term is the \emph{Allee effect},  in the presence of which, unlike in the logistic case,
the zero-$M$ solution is stable. If $M$ is small initially,
it is attracted to the vanishing value; if large, it is attracted
to the nonzero value. The physical origin of the Allee effect is
the possible increase of survival fitness as a function of population
size for low values of the latter. Existence of other members of the
species may induce individuals to live longer whereas low densities
may, through loneliness, lead to extinction. There is a great deal of evidence for such
an effect in nature \cite{ALLEE, ALLEE_1} and there have been recent reports \cite{KK,CLERC} of theoretical work addressing the effect.\\

This paper is set out as follows. The Allee effect is described  by adding cubic terms to the logistic dependence and the model is displayed in section 2, both in dimensioned and dimensionless forms. The former is important to understand the connection of the parameters to quantities observed in nature, while the latter facilitates mathematical analysis. We also define two quantities, $\alpha$ and $\chi$, important to our later development.  The former is central to the classification of regimes of behavior and is directly associated with the Allee phenomenon. We present a linear stability analysis of the steady state solutions in section 3 and argue the importance of the threshold $\alpha=0.5$ on the basis of a Ginzburg-Landau discussion. Restricting attention to the interesting case when $\chi$ exceeds a critical value $\chi_c$, we carry out numerical solutions of the model equations in section 4 for the simple case of spatially constant $\alpha$, and in section 5 for the richer case of spatially modulated $\alpha$. Conclusions appear in section 6.\\
  \indent

\section{The model}
Our starting set of equations is, instead of (1),
\begin{eqnarray}{\label{N_e1}}
\frac{\partial M_S}{dT}&=&bM^2-cM_S-\frac{M_SM^2}{K}-aM_SM_I+D\frac {\partial^2M_S}
     {\partial X^2}, \nonumber \\
\frac{\partial M_I}{dT}&=&-cM_I-\frac{M_IM^2}{K}+aM_SM_I+D\frac {\partial^2M_I}
     {\partial X^2},\nonumber \\
\frac{\partial M}{\partial T}&=&-cM+bM^2-\frac{M^3}{K}+D\frac {\partial^2M}{\partial X^2}.
\end{eqnarray}
The birth term is quadratic in the mouse density and the environment population term is cubic. This is in
contrast to the Fisher case (1) where these terms are linear and quadratic respectively. Needless to say,
using a quadratic versus linear birth rate has nothing to do with sexual versus asexual reproduction. Surely,
the mice do reproduce sexually. The effective power to be used in the equations is the result of a variety
of factors influencing one another including the probability of encounter of mates. The appropriate power of
the density can be
determined only phenomenologically,  after the fact. Similar considerations apply to the quadratic versus cubic
environment term. The important distinguishing feature of the Allee case is the possibility of extinction
that small enough populations have to face as a result of the nonlinearity, and mathematically speaking, the
existence of two (rather than one) stable fixed points.

The simplest way to analyze the steady states of the system is to first look at the solutions of $M$
from the last of Eq. (\ref{N_e1}). There are three possible homogeneous steady state solutions
for $M$: $M_0=0, M_{mx}=(bK+\sqrt{b^2K^2-4cK})/2$ and $M_{mn}=(bK-\sqrt {b^2K^2-4cK})/2$. By performing
the linear stability analysis of these solutions, we find that $M_0$ and $M_{mx}$ are stable while
$M_{mn}$ is unstable. However, it is important to note that if $b^2<4c/k$, there is only one real
solution corresponding to $M=0$. It is convenient to reduce Eq. (\ref{N_e1}) to dimensionless form by performing
the following substitutions:
\begin{eqnarray}{\label{setings}}
   x&=&X\sqrt{M_{mx}^2/KD}, \nonumber \\
   t&=&M_{mx}^2T/K,\nonumber \\
   m&=&M/M_{mx},\nonumber \\
   m_s&=&M_S/M_{mx},\nonumber \\
   m_i&=&M_I/M_{mx}.
\end{eqnarray}
We introduce two new quantities that will prove important in the subsequent analysis:
\begin{eqnarray}
\alpha&=&\frac{M_{mn}}{M_{mx}}=\frac{bK-\sqrt {b^2K^2-4cK}}{bK+\sqrt{b^2K^2-4cK}}, \\
\chi&=&(1+\alpha)\frac{a}{b}.
\label{defax}
\end{eqnarray}
Using $b/M_{mx}=(\alpha+1)/K$ and
  $c/M_{mx}^2 =\alpha/K$, we get the following transformed dimensionless equation set.
   \begin{eqnarray}{\label{N_e3}}
\frac{\partial m_s}{\partial t}&=& (\alpha+1)m^2-\alpha
m_s-m_sm^2-\chi m_sm_i+\frac{\partial^2m_s}{\partial
x^2},\nonumber \\
\frac{\partial m_i}{\partial t}&=&-\alpha m_i-m_im^2+\chi
  m_sm_i+\frac{\partial^2m_i}{\partial x^2},\nonumber \\
\frac{\partial m}{\partial t}&=&m(1-m)(m-\alpha)+\frac{\partial^2m}
    {\partial x^2}.
\end{eqnarray}

\section{Linear Stability Analysis}

Given that the definition of $\alpha$ in Eq. (\ref{defax}) as a mouse density ratio excludes consideration of values of $\alpha$ that are not real,  the only steady values for  $m$ are: 0, $\alpha$ and 1, and there is only one steady stable state for the total population,
corresponding to extinction. The solutions $m=0, 1$ are stable while $m=\alpha$ is unstable.
The crucial consequence of the Nagumo term (cubic nonlinearity) is that
the solution $m=0$ is stable whereas in the Fisher case it is necessarily unstable. Thus, in the presence of Allee effects, it is possible that the mouse population may vanish completely.
Since the system in its steady state adopts the values of  the stable solutions,
we need to analyze the mouse density when $m=0$ or $m=1$.

The case $m=0$ is trivial. When $m=1$, we have only two valid solutions for  $(m_s,m_i)$,
$(1,0)$ and $((\alpha+1)/\chi,(\chi-1-\alpha)/\chi)$. We can
observe, by performing a linear stability analysis, that when
the second solution adopts negative values, it is unstable while the
first one is stable. When $\chi=1+\alpha$, there is a transition,
the first solution becoming unstable and  the second becoming positive
definite and stable. Thus,  $1+\alpha$ represents the critical value of $\chi$:
$$\chi_c=1+ \alpha.$$

The results can be summarized as follows:
\begin{equation}
(m_s,m_i)=\left \{ \begin{array}{lll}
(0,0), (1,0)& \mbox{\ \ if\
\ }&\alpha > \chi-1\\
&&\\
(0,0),&&\\
((\alpha+1)/\chi,(\chi-1-\alpha)/\chi)& \mbox{\ \ if\
\ }&\alpha \leq \chi-1\\
\label{stable}
\end{array}
\right.
\end{equation}

These comments about the fixed points shed light on the main issue of interest in
this paper, viz. the steady state \emph{spatial density profiles}
of mice. Here, we would like to stress that in the presence of Allee effects, the infected mouse
density vanishes for $a<b$. In their absence on the other hand, when the
Fisher equation is appropriate \cite{AK}, the condition
involves not only $a$ and $b$ but also $c$ and $K$. As $\chi$ crosses the value $\chi_c$, there is a change in the
stable character of the solutions. The system
shows an \emph{imperfect pitchfork bifurcation}: the number of fixed points is two
for $\chi<\chi_c$ and three for $\chi> \chi_c$. The bifurcation is imperfect as the system
under study is not symmetric under reflection, i.e., $m_s\rightarrow -m_s$ and $m_i \rightarrow-m_i$.
By contrast, the Fisher modeling of the Hantavirus epidemic
shows a \emph{transcritical} bifurcation \cite{AK}.\\

Our study of this system shows that  $\alpha=0.5$ marks an interesting threshold that separates two regimes corresponding to larger or smaller values of $\alpha$. To see this, recast Eq.(\ref{N_e3}) in the Ginzburg-Landau form, $$\frac{\partial m}{\partial t}=-\frac{\delta{\mathcal{F}}}{\delta
m},$$ where the functional representing the free energy density is given by
\begin{eqnarray}
\mathcal{F}(x,t)&=&\int dx\left[\frac{1}{2}\left(\frac{\partial m}{\partial x}\right)^2
  +F(m)\right]\\
  F(m)&=&-\int_{0}^{m}f(m^{\prime})
 dm^{\prime}.
 \end{eqnarray}
 The dynamics is determined by
 ${\mathcal{F}}$ in that the system evolves such that $\frac{d{\mathcal{F}}}{dt}\le 0$. The
expression for
${\mathcal{F}}$ contains two terms. The first term involving the derivative tries to minimize
${\mathcal{F}}$ by minimizing the
fluctuations in the density. The minima of second term, i.e., $F(m)$ are
governed by the reaction term, $f(m)=m(1-m)(m-\alpha)$ and so we have,
$$F(m)=m^4/4+\alpha m^2/2-(1+\alpha) m^3/3.$$ The significance of the threshold value $\alpha=0.5$ is that $F(0)=F(1)$,
 i.e., solutions corresponding to both $m=0$ and $m=1$ are \emph{equally} stable.
When $\alpha<0.5,$ the solution $m=1$ is relatively more stable than the solution $m=0$. The situation is reversed for
$\alpha>0.5$.

\section{Homogeneous Environment (constant $\alpha$)}

Because infected
mouse density is always zero irrespective of values of $\alpha$  in the subcritical regime $\chi<\chi_c$,  we focus our attention below only
on steady state mouse density profiles for the other, more interesting, supercritical regime, $\chi>\chi_c$. In this case, there is a possibility of getting nonzero $m_i$ and the steady state depends
on the parameter $\alpha$ as well. In order to find the spatial mouse pattern,
we solve Eqs.(\ref{N_e3}) numerically with a given initial spatial distribution of susceptible and infected mouse density.

We consider a bounded domain
where $\chi=2$ everywhere  ( supercritical case since $\alpha<1$) and use reflective boundary
conditions. We analyze various values of $\alpha$ and spatially varying initial densities. Except for   $\alpha= 0.5$,  we find that the system
evolves towards a homogeneous steady state, with values corresponding to the steady solutions.
If we take $\alpha<0.5$ and  $m_s(x)=m_i(x)=A |\cos(2\pi \omega x)|$, a periodically modulated initial condition for the initial population, we
observe that the system evolves to a steady state characterized by
a homogeneous solution, with  $m=1$ and a nonzero value for the infected mouse density,
$m_i=(\chi-1-\alpha)/\chi$, in the entire spatial domain. In other words, the system evolves towards the stable
state corresponding to $m=1$,  and \emph{not} towards the other stable state for $m=0$.
Such a possibility of choice between attractors is a feature associated with  the Allee effect. It does not arise in the earlier description of Hantavirus spread \cite{AK} via Fisher-like equations because there there is, in that description, only one attractor. If, on the contrary, $\alpha >0.5,$
the total mouse density $m$ vanishes completely in the whole spatial domain. This is, in fact, an Allee effect.
The observed effect is in contrast to the result obtained using Fisher-like equation
where population can never go to zero because that would correspond to an unstable state.

The case  $\alpha=0.5$ presents a much richer and complex  behavior of the system. This situation is however, not of biological interest because it corresponds to a set of parameters of measure zero. Although we have obtained some interesting results for that case, we do not display them in detail because of their lack of relevance to observational matters and comment on them only in passing in the Conclusion.

\section{Inhomogeneous environment (spatially varying $\alpha$)}
Noteworthy effects appear when Allee effects combine with a spatially varying $\alpha$. To study them, we maintain the relationship $\alpha \leq \chi-1$ throughout the extension of the whole domain, so that we have a supercritical solution, but assume that the parameters that go into the making of $\alpha$, viz., the birth parameter $b$, the death rate $c$, and the environment parameter $K$ (see Eqs. (\ref{defax})), are spatially varying and consequently introduce a corresponding modulation into
$\alpha$.

To understand transparently the physical meaning of a postulated modulation of $\alpha$ as we take below, it is helpful to consider, only for illustration purposes, the case when the quantity $4c/b^2K$ is small, and to expand the $\alpha$ expression in Eq. (\ref{defax}) in powers of that quantity. Then we get
\begin{equation}
\alpha\approx\frac{4c}{b^2K}.
\label{alpha2}
\end{equation}
Any inhomogeneity in the environment, represented by a spatial variation of $c$, $b$ or $k$ would be reflected in a spatial modulation of $\alpha$. A larger death rate, a smaller environmental parameter or a smaller birth rate would cause a decrease in $\alpha.$

\begin{figure}
\includegraphics[width=9cm]{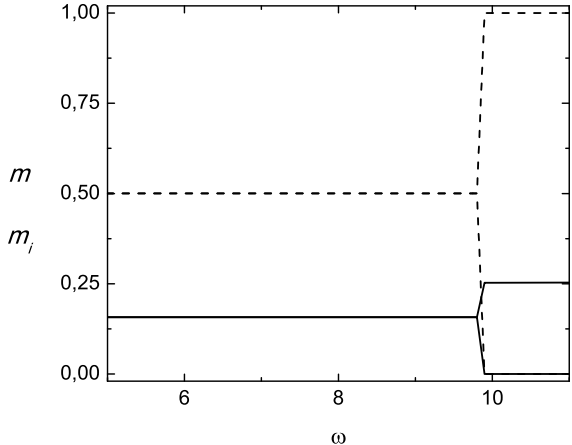}
\caption{Mean values of the total (solid) and infected (dashed) population as a function of $\omega$, the wavenumber that describes the modulation of $\alpha(x)$}.
\label{struca1}
\end{figure}
We have analyzed the effect of such a spatial variation in two forms: first by considering a sinusoidal modulation with a characteristic
wave number and then by considering a less regular behavior of $\alpha$.
First we  take $\alpha(x)=0.5+B \sin(2\pi\omega x)$, with $B<0.5$ so we stay within the region where $\chi>\chi_c$ and avoid $\alpha<0$. Unexpected behavior emerges. The system undergoes a transition as the wave number $\omega$ of the modulation of $\alpha$ crosses a given critical value, $\omega_c\approx 10$.
When $\omega<\omega_c,$ the modulation of $\alpha $ induces a modulation in the population. This occurs around $0.5$ for the total population and around $\frac{\chi-1-\alpha}{2\chi}$ for the infected population. But as $\omega$ exceeds $\omega_c$,  despite the modulation of $\alpha$, the total population of mice adopts a homogeneous profile with two different values. There  are two branches, each one corresponding to the steady solutions shown in Eq.(\ref{stable}) for $\omega>\omega_c$.
Together with the absence of modulation in the total population we observe that the infected population survives throughout the whole domain. While the total population is homogeneous, both the infected and susceptible population are oscillatory in space and take values different from those previously calculated.

\begin{figure}
\includegraphics[width=9cm]{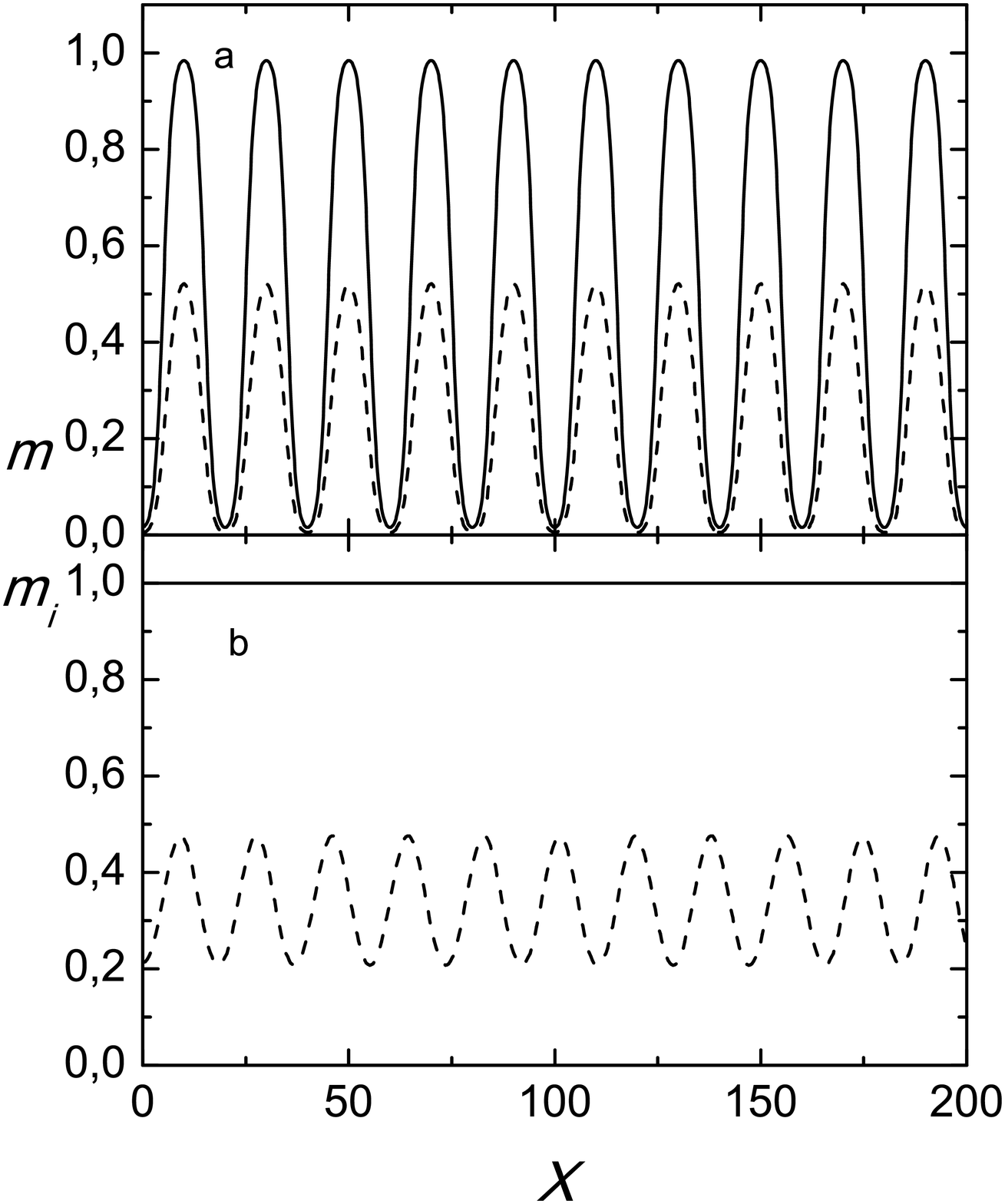}
\caption{Spatial profile of the total (solid) and infected (dashed) mouse population for (a) $\omega<\omega_c$ and  (b) $\omega>\omega_c$ (only the non-null branch is shown). Space is plotted on the x-axis in arbitrary units.}
\label{struca2}
\end{figure}

Fig \ref{struca1} displays graphically the results discussed above. The mean value of the total and infected population is plotted as a function $\omega$, the wavenumber that describes the modulation of $\alpha(x)$. A bifurcation is seen at $\omega_c\approx 10$: the system undergoes a sudden change of behavior. The total population adopts a homogeneous profile while the infected population now survives even in regions where $\alpha(x)> 0.5$. This can be observed in the two plots of Fig. \ref{struca2}, where the spatial profile of both the total and infected populations are shown for values of $\omega$ above and below the value $\omega_c.$

\begin{figure}
\includegraphics[width=9cm]{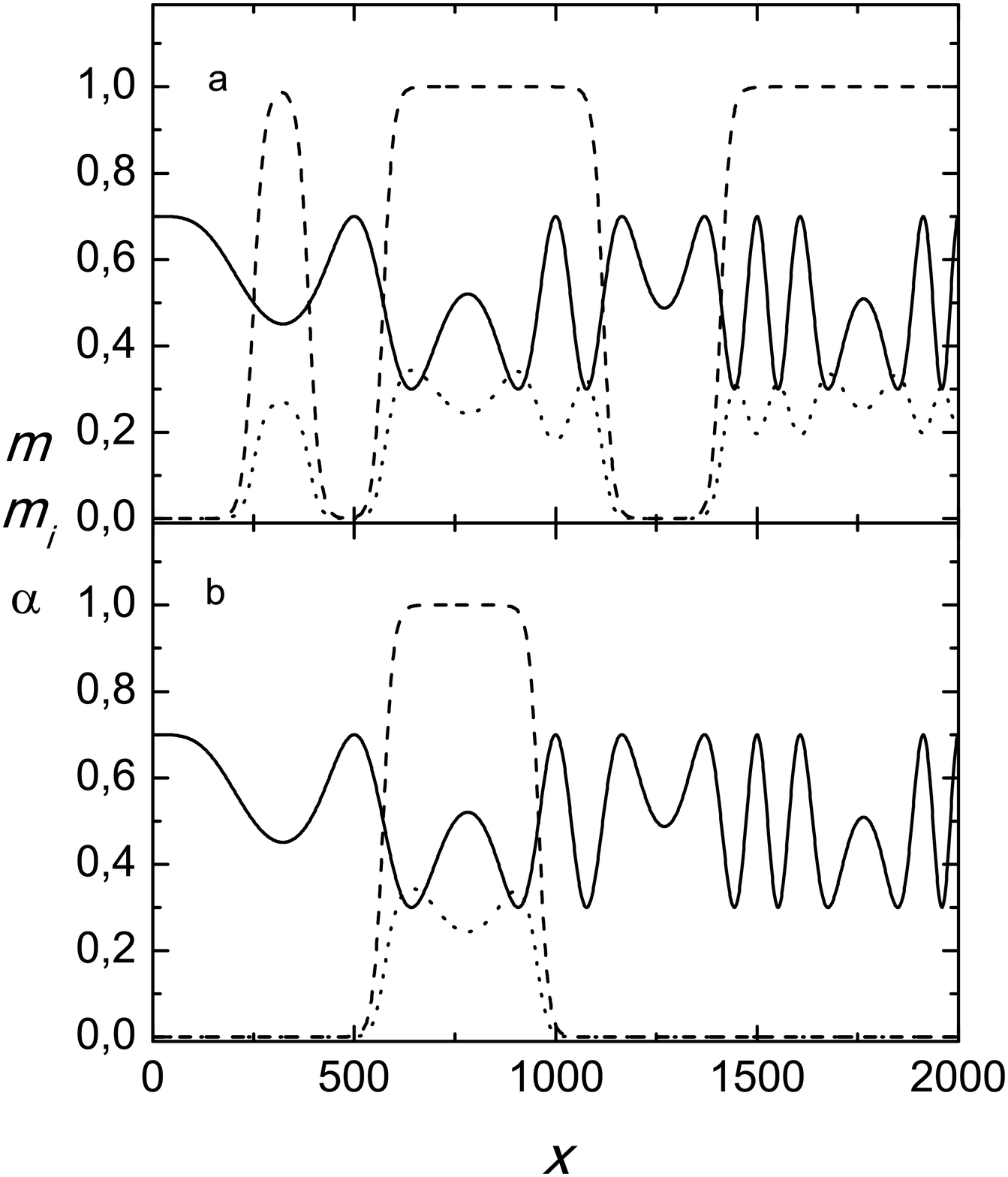}
\caption{Spatial behavior of the total and infected population as $\alpha$ (also displayed) varies in space. Each plot corresponds to a different branch. Depicted are $\alpha$ (solid line), the total mouse population $m$ (dashed line) and the infected mouse population $m_i$(dotted line). Space is plotted on the x-axis in arbitrary units.}
\label{struca3}
\end{figure}

Next we provide an example of a more realistic behavior of the environment that allows us to show that this effect survives even when the modulation is not regular, provided regions exist with oscillations of wavenumber over the critical value.
In Fig \ref{struca3} we plot a situation presenting an irregular profile for $\alpha$ (also displayed in the figure)and the solutions for both branches.

\section{Conclusions}
In summary, we have studied what consequences the possible presence of an Allee phenomenon might have in a population of mice carrying the
Hantavirus epidemic. The Allee features have been incorporated though a Nagumo, i.e. cubic, term in the nonlinearity. We
have analytically solved and examined the fixed points of the problem without diffusion, and numerically
investigated the steady state profiles in the presence of diffusion. Our findings are that, first, as expected, in the
presence of the Allee phenomenon, the mouse population can vanish completely. This is not possible when
the underlying equation is of the type used earlier \cite{AK}. We also observe a dependence from initial conditions
that are not present in the earlier analysis. These effects stem from the existence of two, rather than one,
stable solutions. Formally stated, under the effect of the Allee phenomenon, the system exhibits an
\emph{imperfect pitchfork bifurcation} instead of the transcritical bifurcation observed
in the earlier Fisher case. The most relevant result is the one obtained by introducing environmental spatial inhomogeneities.
The observed effect is better observed in the most abstract case, when the spatial modulation is sinusoidal. We have found that there is a bifurcation in the behavior of the system. The nature of this bifurcation is more evident when calculating the mean value of the population densities.
We have shown the existence of a critical value of the spatial modulation wavenumber $\omega$ at which the behavior of the systems completely changes, displaying bistable behavior that depends on the initial conditions. Later, by taking an hypothetic general situation, we have shown how this effect operates.

For the sake of completeness, we describe briefly the analysis for the physically (observationally) unimportant but mathematically interesting case when $\alpha$ \emph{exactly} equals 0.5. The system behaves in a rather complex way.
Again, we  start with initial conditions of the form $m_s(x)=m_i(x)=A |\cos(2\pi \omega x)|$.
The interesting  feature is that now both steady states are equally stable. The basin of attraction of each state
is such that  the amplitude of the initial modulated condition,  plays a relevant role.
There are three different regimes characterized by 1) the system reaching the homogeneous state
$m=1$ and $m_i=(\chi-1-\alpha)/\chi$ in the whole spatial domain, 2) the total mouse density $m$
vanishing completely in the whole spatial domain, 3) a steady state being characterized by spatial periodic patterns
with $m$ oscillating periodically between 0 and 1. We also observe a similar kind of
oscillation for infected mouse density   between $m_i=0$ and $m_i=(\chi-1-\alpha)/\chi$.
The two first cases are like those encountered already in our analysis above for $\alpha>0.5$, while the third is exclusive to $\alpha=0.5$
As $A$, the amplitude of the initial condition, decreases the system goes from case 1 to case 2 but with an intermediate regime corresponding to case 3. For $\alpha=0.5$, we note the existence of two critical values of initial
conditions at which we observe a transition from  oscillating periodic
structures to homogeneous structures. The first critical value corresponds to a transition
from oscillating structure to a homogeneous pattern with $m=0$.
\begin{figure}
\includegraphics[width=9cm]{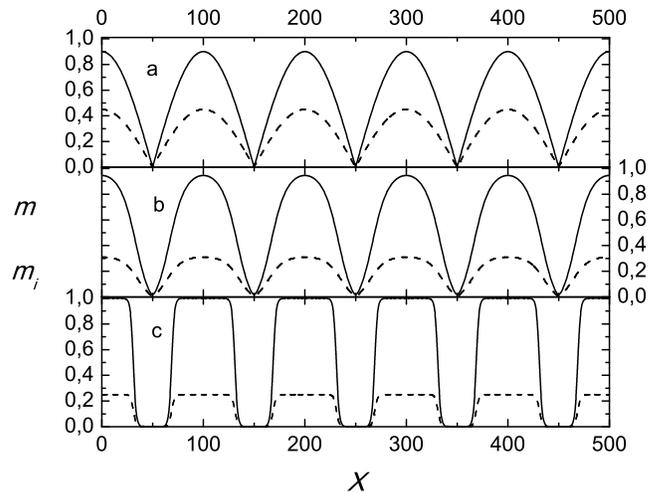}
\caption{The evolution of the initial density profile (a) for $m$(dashed line) and $m_i$(solid line)
 towards the steady state (c) for $\alpha=0.5$ and $a=0.45$. Transient patterns are shown in (b). Here x-axis is the space and y-axis corresponds to density. Units are arbitrary.}
\label{fig:A_0.45}
\end{figure}
The second critical value
corresponds to transition from the oscillating pattern to the homogeneous one with $m=1$. Thus, in
the the presence of Allee effects, the mouse density depends not only on system parameters but also on the \emph{initial distribution}, a dependence that is quite impossible in Fisher equation treatments \cite{AK}.

Whether rodents of the kind we are describing do or do not exhibit these various Allee
effects we have described is a question for the field biologist to pursue. We hope that the interesting consequences that we
have predicted theoretically in this paper will stimulate observational work in this direction.
Work related to the present study (but quite different in spirit as well as detail) may be
found in recent papers by Kenkre and Kuperman \cite{KK} on bacteria in a Petri dish and more recently in the
extensive studies of Clerc et al. \cite{CLERC} who report a number of insights into Allee effects on pattern
formation. Some of the features of the present work that distinguish it from those others is the existence of
infected and susceptible subclasses in the mouse population, from the mathematical point of view the corresponding
bifurcations that occur in the system, and from the physical point of view the relevance of the study to the spread
of epidemics and formation of refugia. In a future publication we will stress a more realistic
2-d depiction of the refugia.

This work was supported in part by the NSF under
grant no. INT-0336343 and by NSF/NIH Ecology of Infectious Diseases under grant no. EF-0326757.

\vspace{5cm}

\end{document}